\documentclass[journal=jacsat,manuscript=article]{achemso}

\usepackage[version=3]{mhchem} 
\usepackage{graphicx}
\usepackage[qm]{qcircuit}
\usepackage{subcaption}
\usepackage{xcolor}
\usepackage{pdfpages}
\usepackage{algpseudocode} 
\usepackage{siunitx}
\usepackage[section]{placeins}
\usepackage{braket}
\usepackage[ruled,linesnumbered,lined, noend]{algorithm2e}
\SetKwProg{Fn}{}{:}{}
\newcommand{\data}{data} 

\usepackage{appendix}

\author{César Feniou}
\affiliation{Sorbonne Universit\'e, LCT, UMR 7616 CNRS, 75005 Paris, France}
\alsoaffiliation{Qubit Pharmaceuticals, Advanced Research Department, 75014 Paris, France}
\author{Olivier Adjoua}
\affiliation{Sorbonne Universit\'e, LCT, UMR 7616 CNRS, 75005 Paris, France}
\author{Baptiste Claudon}
\affiliation{Sorbonne Universit\'e, LCT, UMR 7616 CNRS, 75005 Paris, France}
\alsoaffiliation{Qubit Pharmaceuticals, Advanced Research Department, 75014 Paris, France}
\author{Julien Zylberman}
\affiliation{Sorbonne Universit\'e, LERMA, UMR 8112 CNRS, Paris, 75005 Paris, France}
\author{Emmanuel Giner}
\affiliation{Sorbonne Universit\'e, LCT, UMR 7616 CNRS, 75005 Paris, France}
\author{Jean-Philip Piquemal}
\affiliation{Sorbonne Universit\'e, LCT, UMR 7616 CNRS, 75005 Paris, France}
\alsoaffiliation{Qubit Pharmaceuticals, Advanced Research Department, 75014 Paris, France}
\email{jean-philip.piquemal@sorbonne-universite.fr}

\title[An \textsf{achemso} demo]
  {Sparse Quantum State Preparation for Strongly Correlated Systems}

\abbreviations{IR,NMR,UV}
\keywords{American Chemical Society, \LaTeX}

\begin{document}


\begin{abstract}
Quantum Computing allows, in principle, the encoding of the exponentially scaling many-electron wave function onto a linearly scaling qubit register, offering a promising solution to overcome the limitations of traditional quantum chemistry methods. An essential requirement for ground state quantum algorithms to be practical is the initialisation of the qubits to a high-quality approximation of the sought-after ground state. Quantum state preparation enables the generation of approximate eigenstates derived from classical computations, but it is frequently treated as an oracle in quantum information. In this study, we investigate the quantum state preparation of prototypical strongly correlated systems' ground state, up to 28 qubits, using the Hyperion-1 GPU-accelerated state-vector emulator. Various variational and non-variational methods are compared in terms of their circuit depth and classical complexity. Our results indicate that the recently developed Overlap-ADAPT-VQE algorithm offers the most advantageous performance for near-term applications.

\begin{figure}[!htb]
\centering
   \includegraphics[width=0.5\linewidth]{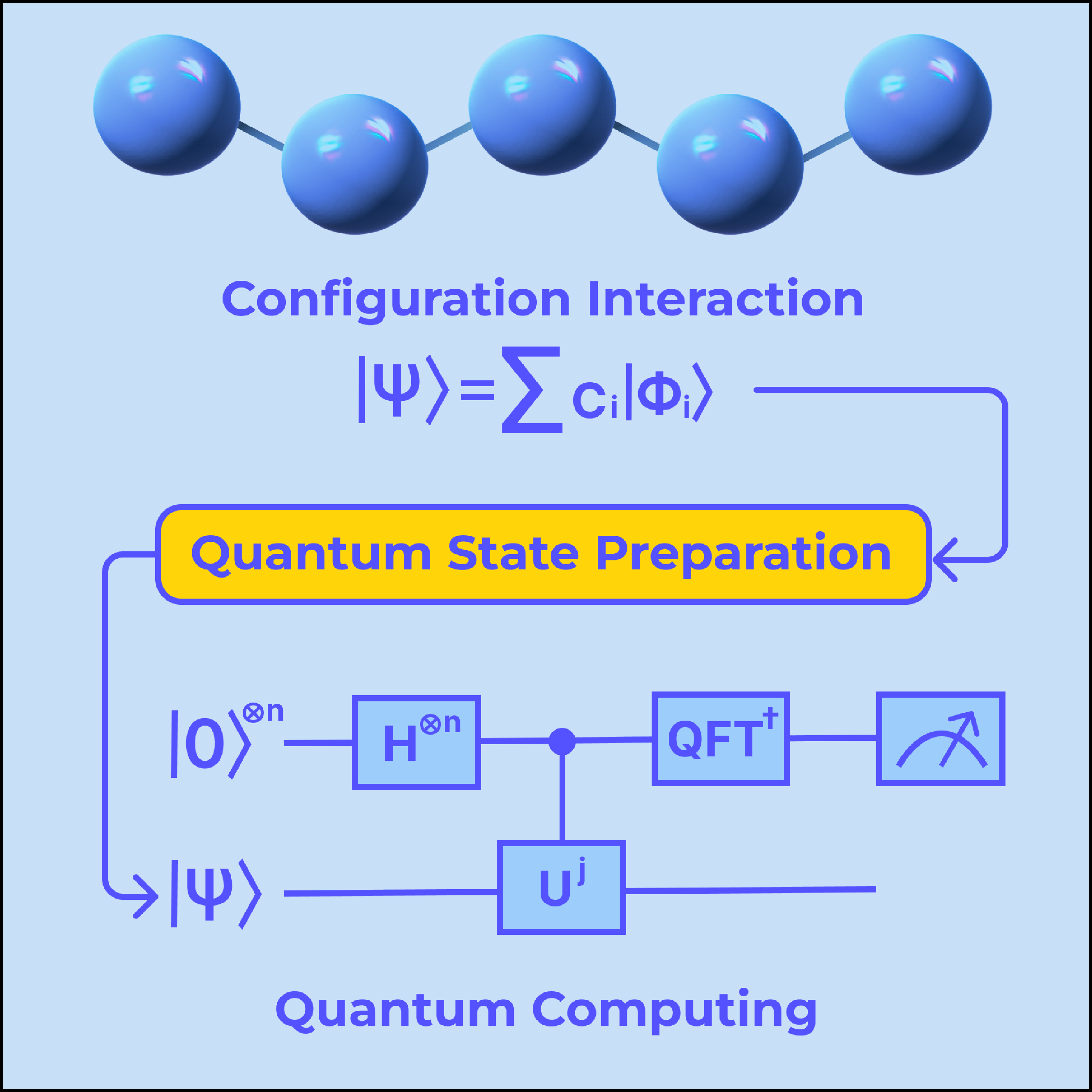}
   
   \label{fig:nodiis} 
\end{figure}



\end{abstract}


Simulating electronic structure on computers is a crucial and challenging endeavor with significant applications in drug discovery and advanced materials design \cite{helgaker2013molecular}. Unfortunately, the full configuration interaction (FCI) method, which provides exact solutions to the electronic time-independent non-relativistic Schrödinger equation in a given basis, requires the diagonalisation of the molecular Hamiltonian to solve the electronic structure problem. The computational cost of performing a FCI calculation therefore increases exponentially as the system size expands. Quantum computers offer a promising solution to overcome this issue by enabling the representation and manipulation of exponentially large electronic wave functions using only a linear number of qubits. To leverage the potential of quantum computers for such tasks, quantum algorithms such as the Variational Quantum Eigensolver\cite{peruzzo2014variational} (VQE) and Quantum Phase Estimation\cite{kitaev1995quantum, nielsen2010quantum} (QPE) have been developed for ground state quantum chemistry calculations\cite{aspuru2005simulated, whitfield2011simulation, tilly2022variational}.

Unfortunately, the limited quantum resources available on current-generation Noisy Intermediate Scale Quantum\cite{preskill2018quantum, bharti2021noisy} (NISQ) devices severely limit the implementation of such methods and their ability to achieve an advantage over their classical counterparts. Hence, there is a need for the simultaneous use of both improved quantum hardware and resource-efficient quantum algorithms. Both VQE and QPE have a significantly reduced computational cost when the initial state, i.e. the starting configuration of qubits, is carefully prepared in order to exhibit high overlap with the ground state of the system under study. Since VQE is an optimization problem, beginning with an initial state that is near the solution naturally enhances the efficiency of the search for the optimal solution. Regarding QPE, the success probability of the algorithm is proportional to the overlap between the initial state and the Hamiltonian ground state being simulated\cite{kitaev1995quantum, nielsen2010quantum}. For a typical chemical quantum simulation, the Hartree-Fock state is usually chosen as initial state for two main reasons: it can be represented on a quantum computer using just a few quantum gates, and it tends to be the predominant Slater determinant for most molecular systems. However, as pointed out in many studies\cite{tubman2018postponing, dalzell2023quantum, tilly2022variational, lee2023evaluating}, a single-determinant initial state might not be accurate enough for quantum computation of strongly correlated systems. One possible strategy to generate more accurate and multireference initial states is to build upon the ability of conventional quantum chemistry methods to produce such wave functions. The subsequent step involves transferring this classical data into the quantum computer to prepare the desired initial state, a process known as \textit{Quantum State Preparation} \cite{plesch2011quantum}, and often treated as 'black-box' or oracle procedure in quantum information. Formally, it involves preparing the state $\ket{\Psi}$ from a set of coefficients $c_p$ such as :
\begin{equation}
\begin{split}
    \ket{\Psi} = \sum_{p \in \{0, 1\}^{n}}c_p\ket{p}, 
    \sum_{p \in \{0, 1\}^{n}}|c_p|^2 = 1.
\end{split}
\end{equation}
Note that an arbitrary quantum state preparation necessitates circuits that scale exponentially with the number of qubits either in size, depth or number of ancilla qubits\cite{grover2002creating,mottonen2004transformation, plesch2011quantum}. Consequently, this initialisation step could potentially dominate the overall computational cost of the quantum algorithm, leading to the loss of the anticipated advantage. A common metric for assessing quantum resource requirements includes the counts of both CNOT gates and single-qubit gates. While the emphasis on CNOT count is typical in NISQ devices, fault-tolerant quantum computing places greater importance on the count of single-qubit gates, which include non-Clifford arbitrary rotations.\\
Configuration interaction\cite{nesbet} (CI) is a post-Hartree-Fock quantum chemistry approach that represents the wave function as a weighted linear combination of Slater determinants. As the size of the FCI space grows exponentially with the number of particles and basis functions, the Selected-CI (SCI) methods have been proposed in order to select automatically only the important contributions in the Hilbert space. 
Among various flavors of SCI, the CI perturbatively selected iteratively (CIPSI)\cite{HurMalRan-1973, three_class_CIPSI} employs an iterative approach in which the most pertinent Slater determinant are dynamically added to the wave function based on an importance derived from perturbation theory (PT). By choosing on-the-fly Slater determinants tailored to the systems being simulated, the SCI ansatz wave function rapidly captures the bulk of correlation effects and is therefore a natural candidate for a suitable multi-reference initial quantum states. 
Formally, the SCI wave function is defined over a subset of Slater determinants within the FCI space, denoted here the variational subset of determinants $\mathcal{V} = \{\ket{\phi_i},i=1,N\}$, and the energy is minimised over this set of parameters 
\begin{equation}
    \begin{split}
        \ket{\Psi_{\text{SCI}}} = \sum_{i \in \mathcal{V} }c_i\ket{\phi_i}, \\
    \{ c_i \} = \text{argmin}\frac{\bra{\Psi_{\text{SCI}}}H\ket{\Psi_{\text{SCI}}}}{\bra{\Psi_{\text{SCI}}}\Psi_{\text{SCI}}\rangle}.
        \end{split} 
\end{equation}
For a given wave function $\ket{\Psi_{\text{SCI}}} $, one can compute the associated second-order perturbed energy as 
\begin{equation}
    E^{(2)} = \sum_{I \notin \mathcal{V}} \frac{\bra{I}H\ket{\Psi_{\text{SCI}}}^2}
     {E(\Psi_{\text{SCI}}) - E(I)},
\end{equation}
where $E(\Psi) $ is the variational energy of a given wave function $\Psi$. 
Therefore, the quantity $E^{(2)}$ is a measure of the remaining contribution to the energy of the Slater determinants not already captured in $\ket{\Psi_{\text{SCI}}}$. In the following calculations, we stop our calculations when $|E^{(2)}| < 10^{-4}$ in order to obtain target states of near FCI quality. 

The process of encoding CI states in qubit registers involves mapping Slater determinants to computational basis states. The Jordan-Wigner fermion-to-qubit transformation\cite{jordan1993paulische} is commonly employed for this mapping. In this transformation, each qubit stores the occupation of a spin-orbital, meaning that each vector in the computational basis corresponds to a Slater determinant. Computational basis states can represent unphysical states, like the $\ket{0}^n$ state, which corresponds to a configuration with zero particles. The Hartree-fock state of an $m$-electron and $n$-spin-orbital system is encoded as $\ket{\Psi_{\text{HF}}} = \ket{1}^m \otimes \ket{0}^{n-m}$, when arranging the spin-orbitals in increasing order of energy. Firstly, let us remark that the SCI target state has a number of variational determinants $M$ much smaller than the dimension of the computational Hilbert space ($M\ll2^n$). It is thus encoded as a \textit{sparse quantum state} of sparsity $M$. Second, more generally, let us remark that it respects multiple fundamental symmetries such as fermionic anticommutation relations, which are reflected in the amplitudes, particle conservation and projected-spin, with only specific computational basis states respecting them. The molecular hamiltonian eigenstates thus belong to a subspace of the n-qubit Hilbert space that is relevant to these symmetries. \\

Having established the properties of the target wave functions, we can now shift our attention to exploring the diverse methods available for quantum state preparation of such targets. One first natural idea was introduced by Gard et. al.\cite{gard2020efficient} and involves fully parameterising the symmetry-preserving subspace with minimal CNOT count. Parameters are then variationally tuned until convergence to the target state which is within this subspace. Assuming convergence, this Symmetry-Preserving-VQE approach allows for an exact state preparation with a maximum CNOT count of three times the dimension of the subspace. 
An established alternative to fixed circuit methods involves adaptive variational algorithms where quantum circuits are grown tailored to the problem being simulated. The Adaptive Derivative-Assembled Pseudo Trotter (ADAPT)-VQE \cite{grimsley2018adapt} belongs to this family of methods, and has become a benchmark technique due to its capability of generating ansatz wave functions that are both highly accurate approximations to the ground state and also more compact in terms of circuit depth than fixed-ansatze approaches. A variant of this approach-- labelled Overlap-ADAPT-VQE \cite{Feniou_2023}-- was recently introduced for quantum state preparation. The Overlap-ADAPT-VQE algorithm iteratively generates a compact approximation of a target wave function through a quasi-greedy procedure that maximises, at each iteration, the overlap of the current iterate with the target. This is achieved by adding on-the-fly, the most relevant unitary operator from a finite-size pool of admissible operators, with a selection criterion based on the gradient of the overlap of the ansatz with the target. Note that whether the Overlap-ADAPT-VQE ansatz evolves within the symmetry-relevant subspace of the target depends on the choice of the operator pool. Pools made of fermionic operators adhere to all the desired symmetries \cite{grimsley2018adapt}, which facilitates the convergence to the target state but comes at the expense of requiring more quantum gates per operator. Conversely, Qubit-Excitation-Based \cite{yordanov2021qubit, ryabinkin2018qubit}, Qubit\cite{tang2021qubit}, or Minimal ZY \cite{tang2021qubit} pools may respect only a subset of these symmetries or none at all, but they are more resource-efficient. The Overlap-ADAPT-VQE workflow and the associated details of the aforementioned operator pools are provided in the Appendix.\\

Despite a rather intuitive structure, variational quantum state preparation algorithms face scalability issues because they rely on heuristic classical optimisation processes that become exponentially complex as the number of qubits grows \cite{PhysRevResearch.5.033225}. Indeed, the optimisation problem is non-convex and plagued by multiple local minima and barren plateaus \cite{mcclean2018barren}, which so far limited the applicability of variational quantum state preparation algorithms to rather simple quantum systems. Recently, significant research attention directed towards non-variational methods for quantum state preparation \cite{araujo2021divide, araujo2023low, zhang2022quantum, sun2023asymptotically}, particularly when the state to be prepared exhibits distinct structural traits or symmetries\cite{zylberman2023efficient, holmes2020efficient, marin2023quantum, quantum_dac, tubman2018postponing, zhang2022quantum, de_Veras_2022}. As a result, non-variational sparse quantum state preparation algorithms offer an interesting alternative for the preparation of SCI wave functions, as they are exact and deterministic. We propose in Table \ref{tab:methods_and_cost}, a non-exhaustive list of sparse quantum state preparation algorithms and associated complexities. 
The 'CVO-QRAM' algorithm developped by De Veras et al.\cite{de_Veras_2022} enables sparse quantum state preparation algorithm that scales linearly with respect to the sparsity of the target, and using a single ancilla qubit. 
We detail in the appendix the correctness of the algorithm and use it in the following simulations.

\begin{table}[H]
    \centering
    \resizebox{\linewidth}{!}{%
    \begin{tabular}{|c|c|c|c|c|}
    \hline
    \textbf{Method} & \textbf{Classical preprocessing} & \textbf{CNOT count} & \textbf{Circuit Depth} &\textbf{Ancillas}\\
    \hline
    CVO-QRAM \cite{de_Veras_2022} & $\mathcal O(M\log M+nM)$ & $\sum_{t=1}^n\mu_t(8t-4)-t_{\text{max}}$ &$\mathcal O(nM)$& 1\\
    \hline
    
    Gleinig et al. \cite{quantum_dac}& $\mathcal O(M^2\log (M)n)$ & $\mathcal O(nM)$&$\mathcal O(nM)$& 0\\
    \hline
    Tubman et al. \cite{tubman2018postponing}&& $\mathcal O(nM)$ &$\mathcal O(nM)$& 1\\
    \hline
    Zhang et al. \cite{Zhang_2022}& $\mathcal O(nM)$& [very large]
    &$\Theta(\log(nM))$& $\mathcal O(nM\log M)$\\
    \hline
    Fomichev et al. \cite{fomichev2023initial}&$\mathcal O(M^3)$ & $\mathcal O(M\log M)$ &$\mathcal O(M\log M)$& $\mathcal O(\log M)$\\
    \hline
    \end{tabular}
    }
    \caption{Performance of common sparse state preparation methods.}
    \subcaption{With respect to the target state, $M$ denotes the sparsity, $n$ the number of qubits involved, $\mu_t$ is the number of determinants with $t$ bits of value $1$ and $t_{\text{max}}$ is the highest value of $t$ with $\mu_t\neq0 $.}
    \label{tab:methods_and_cost}
\end{table}

\begin{figure}[H]
    \centering
    \begin{subfigure}{0.49\textwidth}
        \includegraphics[width=\linewidth]{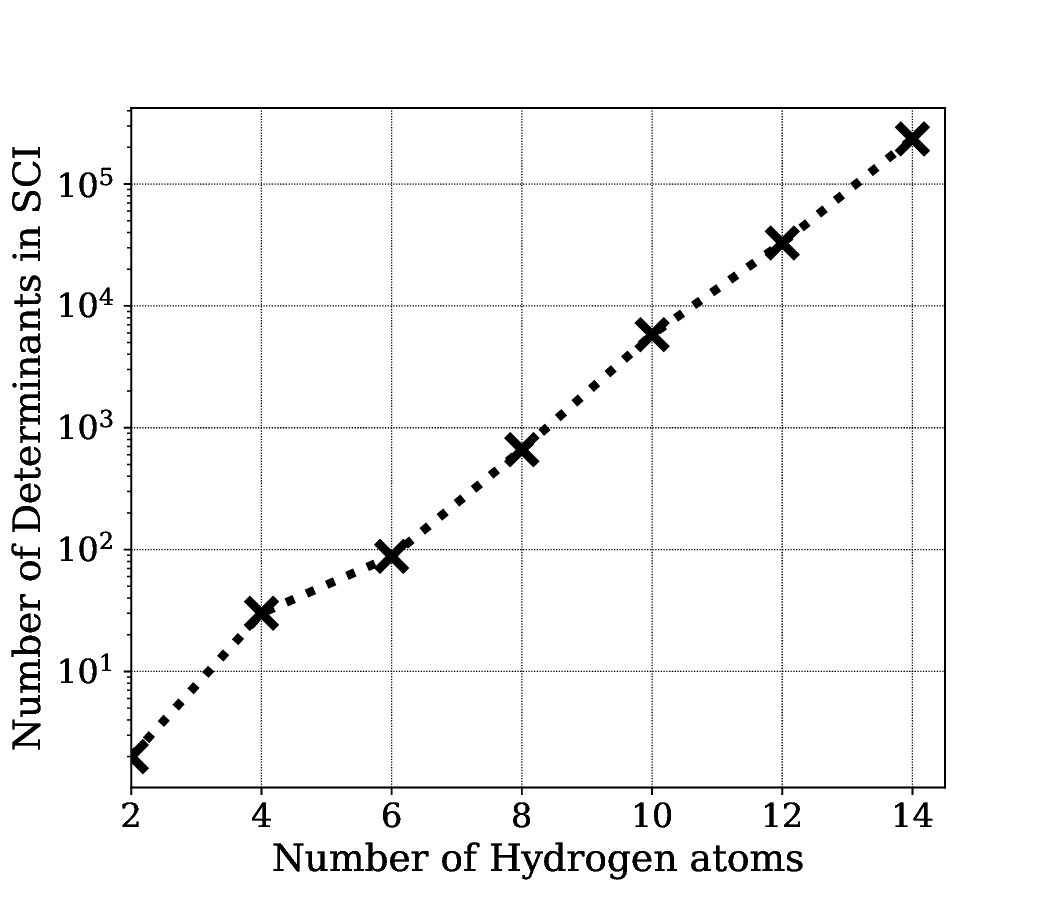}
        \caption{}
        \label{fig:subfig1}
    \end{subfigure}
    \hfill
    \begin{subfigure}{0.49\textwidth}
        \includegraphics[width=\linewidth]{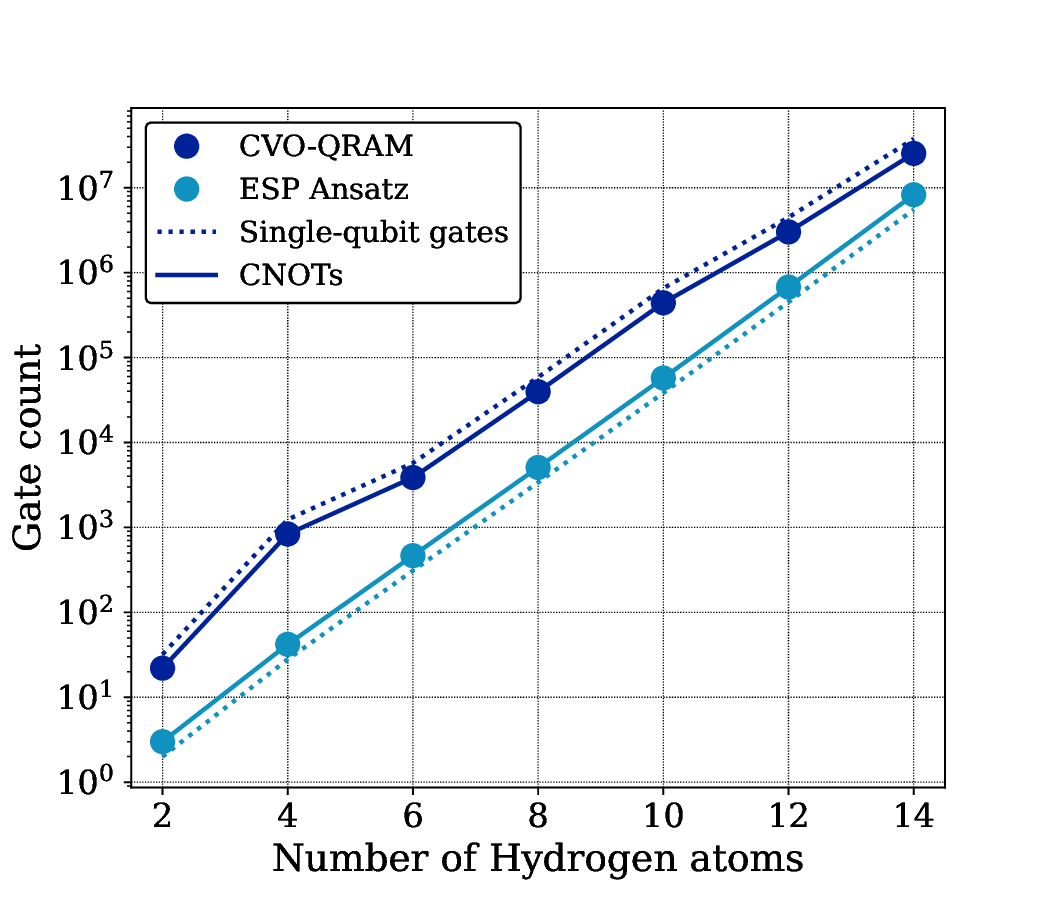}
        \caption{}
        \label{fig:subfig2}
    \end{subfigure}
    \caption{Determinants count (a) and circuit size (b) in the SCI representation of linear H$_n$ chains. The circuit size has been established using CVO-QRAM loader and ESP Ansatz as state preparation method.}
    \label{fig:main}
\end{figure}

The study presented in this Letter aims to compare the aforementioned state-of-the-art variational and non-variational approaches to sparse quantum state preparation on the criteria of circuit size and classical processing complexity. The target states under consideration are accurate approximations of hydrogen chains' ground states, which are prototypical examples of strongly correlated systems in quantum chemistry\cite{10.1063/5.0014928}. The linear hydrogen chains that we consider are stretched with an interatomic distance of $5.0~\text{\AA}$. The target ground states are obtained via CIPSI\cite{HurMalRan-1973, three_class_CIPSI} SCI simulations carried out on the Quantum Package\cite{doi:10.1021/acs.jctc.9b00176} software, using a minimal STO-3G basis set, and up to the challenging case of H$_{14}$, whose quantum encoding requires the significant count of 28 qubits. Figure \ref{fig:subfig1} emphasises, as anticipated, the exponential growth in the number of determinants needed to represent the ground-state of increasingly large hydrogen chains. 

In Figure \ref{fig:subfig2}, we display the circuit size required to construct a quantum circuit that encodes the Full-CI wave functions of hydrogen chains. These quantum circuits are obtained using the CVO-QRAM method and the Efficient Symmetry-Preserving (ESP) ansatz as quantum state preparation algorithms applied to the CIPSI ground state as a target. While the former approach substantially lowers the CNOT count compared to the CVO-QRAM method, it is crucial to emphasise that the ESP Ansatz is built upon certain underlying assumptions. First, the convergence of this optimisation process cannot be guaranteed, giving rise to system-dependence, while CVO-QRAM ensures systematic success. Second, variational state preparation via ESP Ansatz is of exponential complexity with the system size. In either scenario, performing state preparation for such target states presents challenges, with impractical computational cost in one case and large circuit size detrimental to the subsequent quantum post-treatment in the other. 


The CVO-QRAM and ESP quantum state preparation estimated costs suggest that preparing compact approximations of the target state might be a natural option. Notably, variational methods can progressively approach the solution, allowing access to ansatz wave functions of arbitrary precision with flexible circuit depth and classical computational time. Adaptive variational algorithms have notably showcased their effectiveness compared to fixed-ansatz techniques in the realm of approximate ground state preparation. Specifically, the Overlap-ADAPT-VQE has previously been applied to the state preparation of small SCI-wave functions\cite{Feniou_2023}, making it a suitable choice for the forthcoming simulations. 
\begin{figure}[H]
    \centering
    \begin{subfigure}{0.5\textwidth}
            \includegraphics[width=\linewidth]{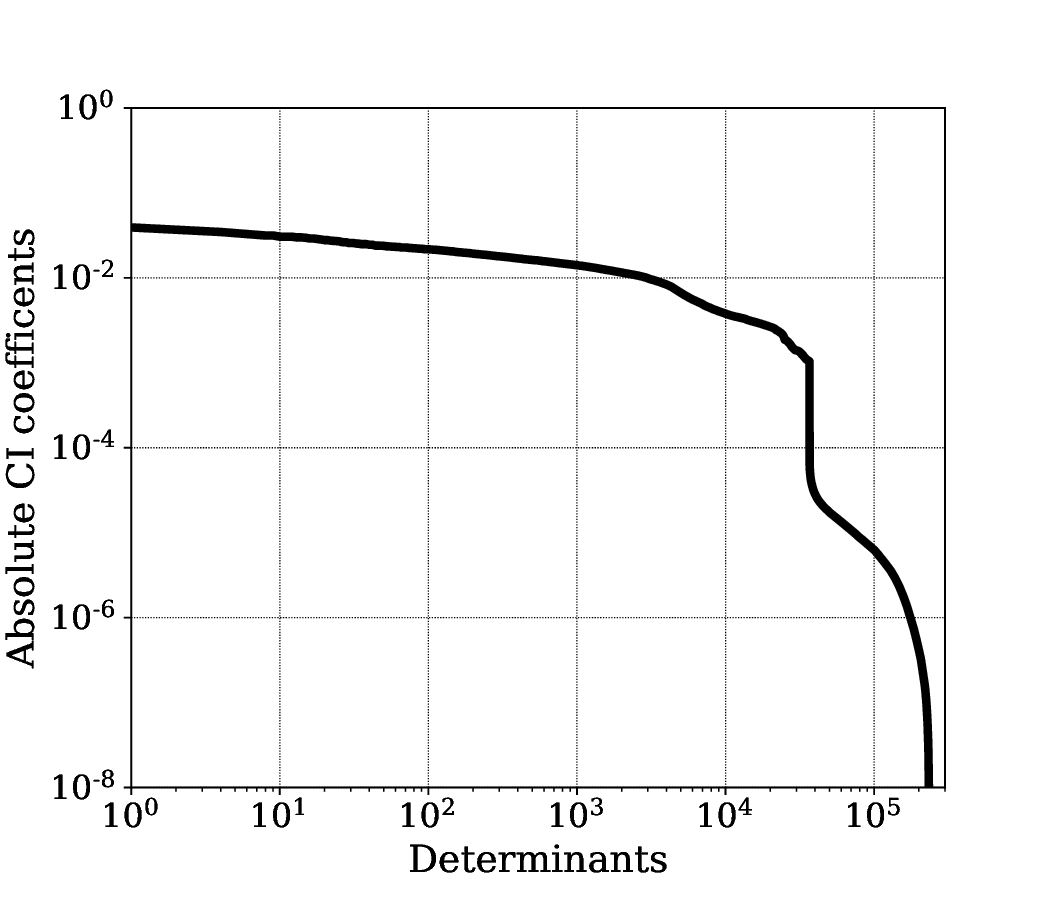}
        \caption{}
        \label{fig:amplitude}
    \end{subfigure}
    \hfill
    \begin{subfigure}{0.48\textwidth}
        \includegraphics[width=\linewidth]{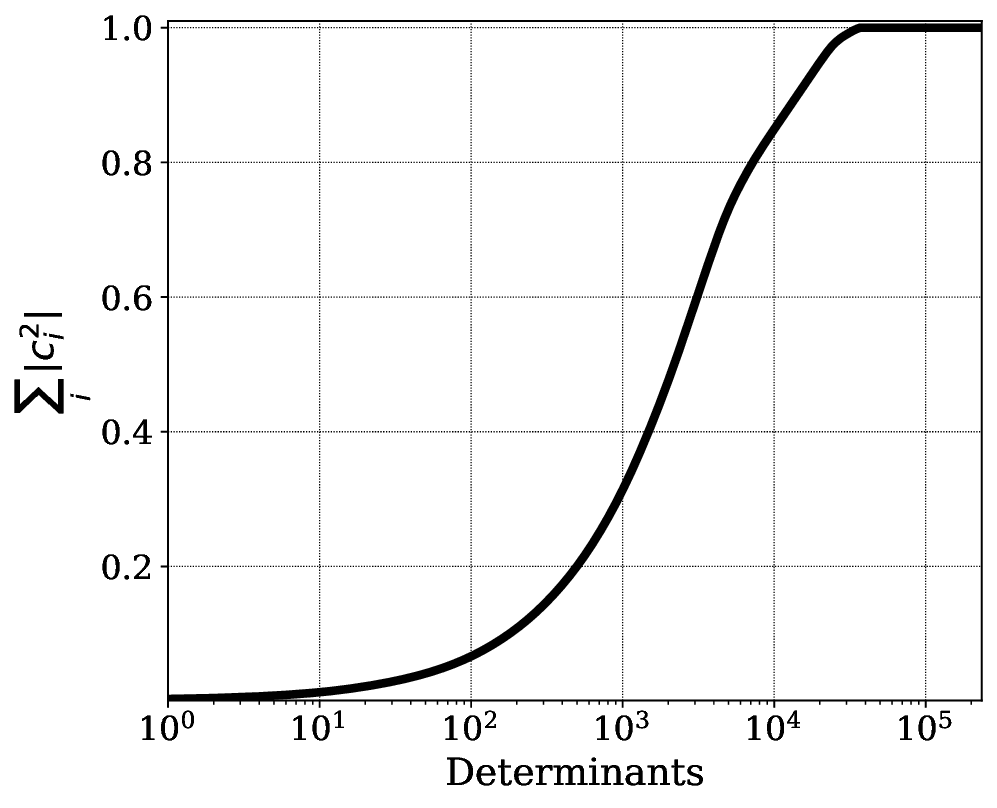}
        \caption{}
        \label{fig:ci_squared}
    \end{subfigure}
    \caption{(a) Absolute CI coefficient of each Slater determinant in the Selected-CI expansion of linear H$_{14}$ ground state wave function. (b) Associated absolute sum of CI squared coefficients.}
    \label{fig:main}
\end{figure}
We carry out Overlap-ADAPT-VQE simulations with the linear H$_{14}$ ground state as target on 28 qubits, and using two common operator pools, namely Qubit-Excitation-Based- and Qubit-pool, whose details are given in appendix. Such large computations are made possible thanks to the use of the in-house GPU-accelerated Hyperion-1 quantum emulator. To ensure a fair comparison with CVO-QRAM, which performs exact state preparation, we prepare a range of approximated states of the linear H$_{14}$ ground state. The approximation involves cropping the smallest coefficients of the target and renormalising, guaranteeing minimal sparsity for an arbitrary high overlap. An alternative being explored involves taking as approximated ground state the variational states achieved during each iteration of CIPSI. These states bear higher physical significance, and have been observed as accurate ansatz-drivers in the context of post-treatment with the VQE\cite{Feniou_2023}. A table comparing the fidelities and the determinant count of these two options is provided in the appendix \ref{sec:target}. 
We display in Figure \ref{fig:amplitude} the absolute weight of each Slater determinant in the CI of the H$_{14}$ linear hydrogen chain ground state. Note that for the H$_{14}$ system, relying on a single determinant initialisation yields at most a modest $4.10^{-2}$ overlap with the ground state. Moreover, Figure. \ref{fig:ci_squared} presents the associated absolute sum of CI squared coefficients, highlighting the need for a multi-determinant initial state. 

\begin{figure}[H]
    \centering
    \begin{subfigure}{0.49\textwidth}
        \includegraphics[width=\linewidth]{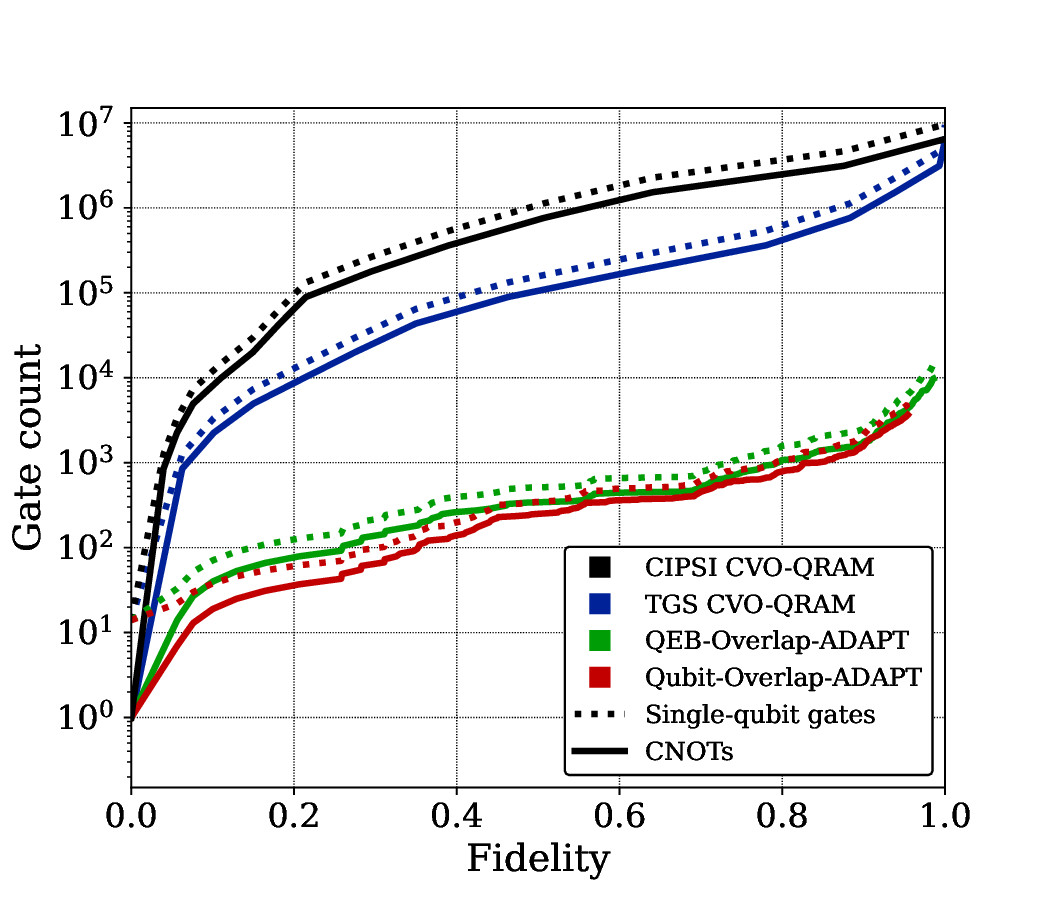}
        \caption{}
        \label{fig:cnot_ovlp}
    \end{subfigure}
    \hfill
    \begin{subfigure}{0.49\textwidth}
        \includegraphics[width=\linewidth]{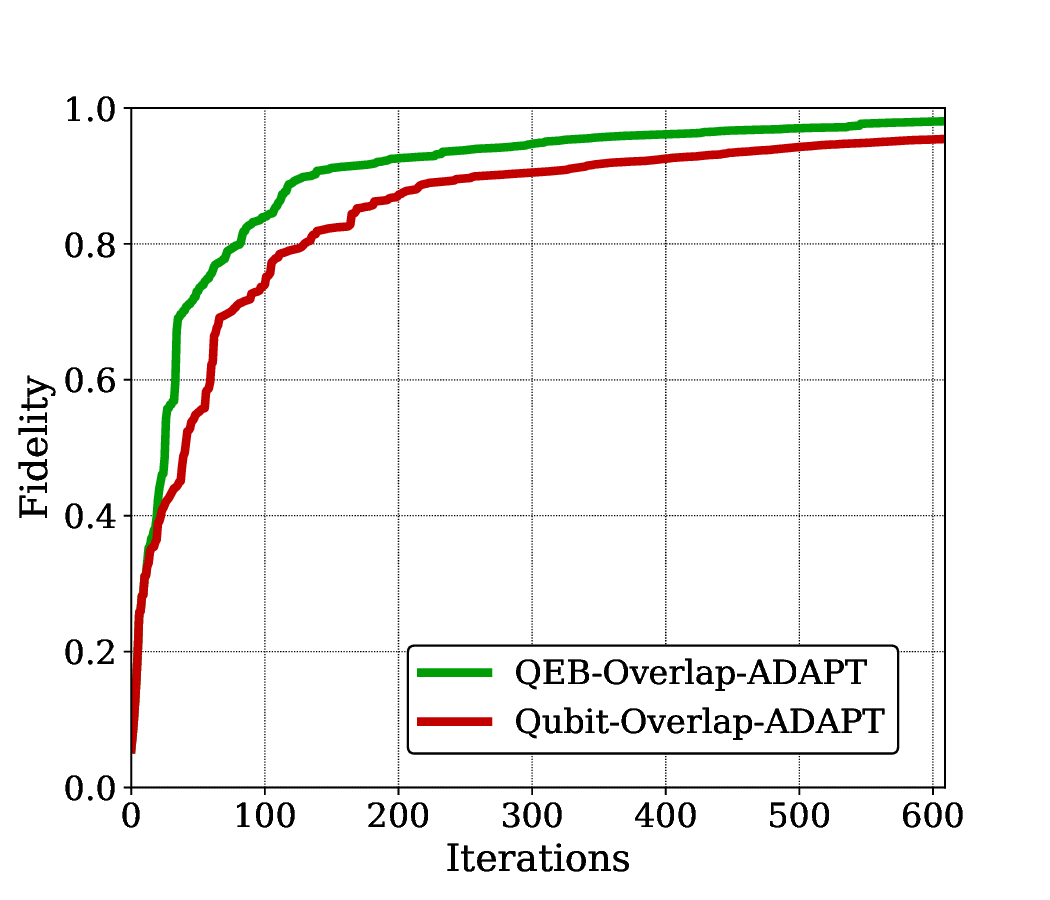}
        \caption{}
        \label{fig:iterates}
    \end{subfigure}
    \caption{(a) Gate-count for approximate state preparation of ground state wave function of linear H$_{14}$ displayed in \ref{fig:main}. The CVO-QRAM algorithm generated states from CIPSI iterates as well as states derived from truncations of the ground state (TGS). The Overlap-ADAPT-VQE approximates the ground state using the QEB- and the Qubit-pool of operators. (b) Fidelity of the Overlap-ADAPT-VQE ansatz over the iterations for the same target.}
    \label{fig:ovlap}
\end{figure}
The resulting gate count of state preparation via CVO-QRAM and Overlap-ADAPT for various levels of approximations are given in Figure \ref{fig:cnot_ovlp}. The adaptive variational approach leads to significantly reduced gate count for any observed level of fidelity. Overlap-ADAPT-VQE seamlessly expands ansätze with fidelity exceeding 95\%, and appears to be resilient to common variational algorithm challenges, such as hostile landscapes with barren plateaus or local minima. This observation is particularly noteworthy considering the strong correlation inherent in the target system. The associated circuit involves no more than a few thousand CNOT gates and single-qubit gates, a highly reasonable resource in the context of the early-fault tolerant era. 
We anticipate the dominance of the CVO-QRAM approach in exactly preparing the target state. Achieving full convergence with the variational approach is unlikely without an exceedingly flexible ansatz involving an impractical number of parameters. This is primarily due to the abundance of determinants with very small weights in the target state. Indeed, Figure. \ref{fig:iterates} shows the convergence profile of the Overlap-ADAPT simulations and illustrates the challenges faced by the ansatz in reaching a fidelity of 1 throughout the iterations. Nevertheless, the majority of studies typically regard an overlap exceeding 0.5 with the ground state as sufficiently high for a QPE initial state. This perspective makes the exact state preparation approach via CVO-QRAM less suitable for these applications. 

The CVO-QRAM process exhibits a superlinear classical complexity with the number of determinants, which are obtained with classical SCI methods that scale exponentially with the system's size. Therefore, the overall state preparation strategy is dominated by the cost of the SCI procedure. When targetting a classically-derived wavefunction, the Overlap-ADAPT-VQE process necessitates classical simulation of a quantum circuit, leading to exponentially scaling memory requirement with the number of qubits and, consequently, with the size of the targeted system. This imposes an upper limit on the qubit count that can be emulated without resorting to approximation or encountering errors. To extend the scalability of variational algorithms on simulators, alternative strategies have been investigated. These include approximated emulations by using Matrix Product State circuit simulators or by partitioning the system and performing successive local optimisations. These efforts have shown promising outcomes, successfully simulating systems of up to around 100 qubits\cite{shang2023towards}. Despite multiple layers of approximation, integrating Overlap-ADAPT-VQE with approximated emulation can be promising in replicating high-fidelity ansätze using compact circuits for larger systems, a focus for future research, particularly for molecular systems with classically intractable convergence. To capitalise on quantum computing memory advantages while maintaining the objective of high-fidelity initial states with compact circuits, an alternative approach is to employ the Overlap-ADAPT-VQE algorithm as a hybrid quantum-classical algorithm. As it necessitates the target wave-function to be stored in a quantum register, executing this method for our problem first involves converting the CIPSI wave-function into a quantum state using CVO-QRAM. Then, an Overlap-ADAPT-VQE ansatz can be grown iteratively to maximise its overlap with the CVO-QRAM prepared state, potentially resulting in a more compact approximation of the latter. The overlap measurement can be conducted using either the swap test or the compute-uncompute approach, both of which have been explored on quantum hardware in prior research within this context (\cite{feniou2023greedy}). Regardless of the chosen approach, let us emphasise that Overlap-ADAPT-VQE not only delivers a precise initial state but presents it in the structure of a parameterised quantum circuit. This circuit can be fine-tuned on the quantum computer, offering adaptability based on the desired post-treatment. This feature proves especially advantageous in the realm of VQE, where flexibility in the ansatz is sought.

By generating a compact quantum circuit that encodes the best classically-derived wave function, we secure an initial quantum state with strong overlap with the sought-after ground state of the system. This initial state preparation will greatly facilitate a following quantum post-treatment, such as QPE, which offers asymptotic advantage in determining the ground state energy compared to classical approaches. \\ 
The key takeaway is that, even for chemical systems of prime hardness\cite{choi2022improving}, quantum state preparation of accurate initial state has proven seamless with Overlap-ADAPT-VQE coupled to a classical ground state approximation procedure. The resulting quantum circuits exhibits an insignificant circuit depth in the context of early-fault-tolerant era. It is worth noting that an entirely different class of procedures conducts state preparation directly within the quantum computer, either through adiabatic state preparation\cite{aspuru2005simulated} or by leveraging ground-state boosting\cite{wang2022state} based only on the information provided by the system's Hamiltonian. While these approaches may seem more scalable than classical state preparation algorithms, they introduce a substantial quantum resource overhead for medium-sized systems, making them inherently more appropriate for the longer term. Further investigation is sought for addressing the regime where the strategy of utilising classically-derived initial states outperforms other initialisation approaches.
In a broader context, sparse quantum state preparation methods are used for a range of applications, including Hamiltonian simulation with either linear combination of unitary operators\cite{childs2012hamiltonian} or qubitization\cite{low2019hamiltonian}, data loading for quantum machine learning\cite{zhao2021smooth} or even solving linear system of equations\cite{harrow2009quantum}. Overlap-ADAPT-VQE may therefore offer a practical approach for any such algorithm involving classical data loading by allowing the use of an operator pool tailored to the target state's characteristics.

\section{Technical Appendix}
\subsection{Quantum state preparation for H$_{10}$ and H$_{12}$ chains}
We display in figure \ref{fig:cnot_ovlp} the gate count for quantum state preparation of the ground state for two additional hydrogen chains, H$_{10}$ and H$_{12}$. Although the Overlap-ADAPT-VQE algorithm demonstrates greater size efficiency in both cases, with an order of magnitude fewer gates compared to the CVO-QRAM in terms of ground-state fidelity, the disparity is less pronounced than in the case of H$_{14}$. The observed difference can be attributed to the lower degree of correlation in shorter hydrogen chains, with the Hartree-Fock state having stronger support and fewer determinants required to accurately approximate the ground state.
\begin{figure}[H]
    \centering
    \begin{subfigure}{0.49\textwidth}
        \includegraphics[width=\linewidth]{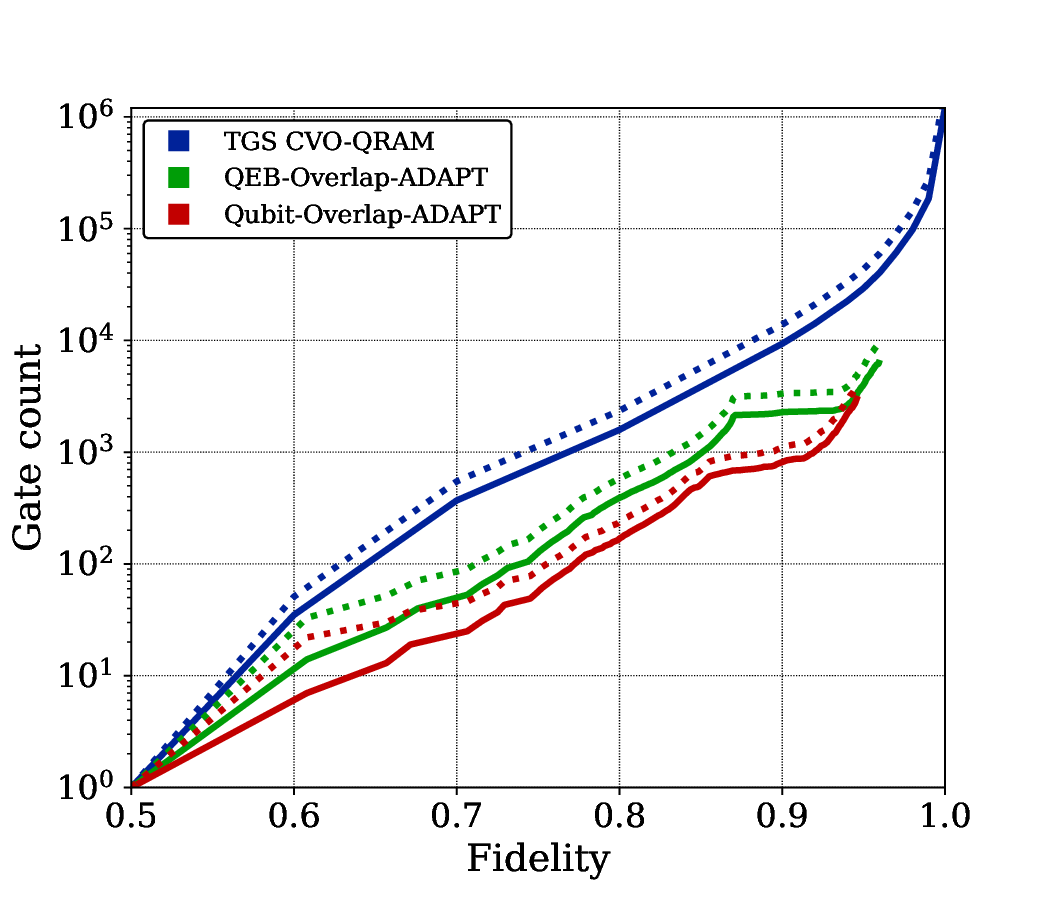}
        \caption{H$_{10}$}
        \label{fig:cnot_ovlp_h10}
    \end{subfigure}
    \hfill
    \begin{subfigure}{0.49\textwidth}
        \includegraphics[width=\linewidth]{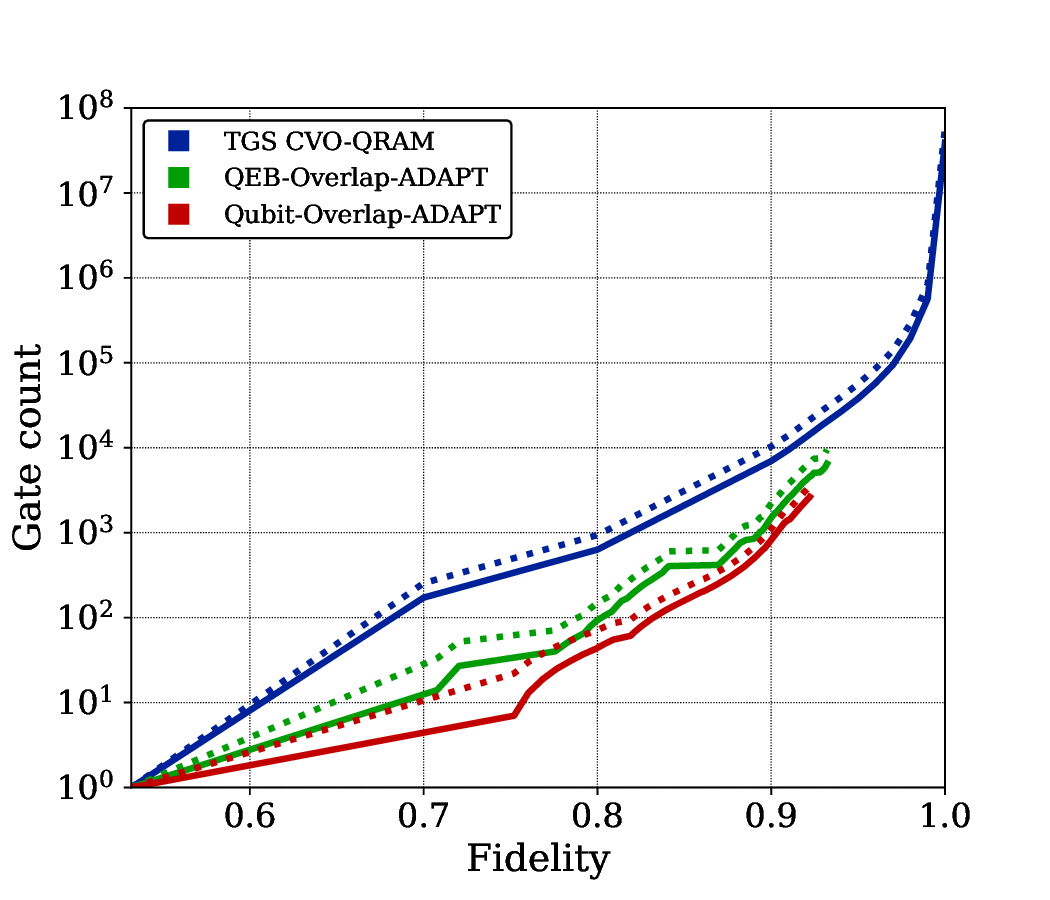}
        \caption{H$_{12}$}
        \label{fig:cnot_ovlp_h10}
    \end{subfigure}
    \caption{Gate-count for approximate state preparation of ground state wave function of linear H$_{10}$ in (a) and H$_{12}$ in (b). The CVO-QRAM algorithm generated states derived from truncations of the CIPSI ground state (TGS). The Overlap-ADAPT-VQE approximates the CIPSI ground state using the QEB- and the Qubit-pool of operators, iterating over 500 steps for each system. The solid lines represent the CNOT counts, while the dashed lines indicate the single-qubit gate counts.}
    \label{fig:cnot_ovlp}
\end{figure}
\subsection{The Hyperion-1 High-Performance State-Vector Emulator}
Hyperion-1 \cite{adjoua2023} is a quantum computing state-vector emulator package dedicated to quantum chemistry. It uses classical hardware and is massively parallel thanks to an efficient multi-GPU (GPU=Graphics Processing Unit) implementation. High performance is ensured by an ensemble of fast custom CUDA sparse linear algebra libraries, which act on the native quantum chemistry algorithms to accelerate Hyperion-1's exact/noiseless simulations. For example, using a single DGX-A100 node (8 X A100 GPUs, 40Gb of memory/GPU), we have been able to carry out a 28 qubits Overlap-ADAPT-VQE simulation producing over 600 iterations within 8 days. The fidelity of the ansatz with respect to the H$_{14}$ target state is plotted with respect to the iterates in Figure. \ref{fig:iterates}. The algorithms examined in this manuscript are classical, and their execution in the state-vector formalism excludes both complete gate decomposition and the simulation of noise related to shots or hardware. Nevertheless, these options are accessible in the software. Further details about Hyperion-1 will be given in a forthcoming publication.

\begin{appendix}\subsection{Double Sparse Quantum State Preparation}\label{sec:cvoqram}

Even though it is unclear how to prepare an arbitrary state of $2^n$ complex amplitudes with a cost which is not exponential with the number of qubits $n$, the problem of loading a number 

$M$ of complex amplitudes was proven to have satisfying solutions scaling linearly with $M$ such as the CVO-QRAM method\cite{de_Veras_2022}. This algorithm consists in loading sequentially the classical data by performing controlled rotations using one ancilla qubit. The algorithm is presented and the correctness is proven recursively as follows.

\begin{algorithm}
  \SetKwFunction{load}{cvoqram}
  \SetKwInOut{Input}{input}\SetKwInOut{Output}{output}
  \Input{data = $\{x_k, p_k\}_{k=0}^{M-1}$}
  \BlankLine
  \Output{$\ket{m} = \sum_{k=0}^{M-1} x_k \ket{p_k}$}
  \BlankLine
  \Fn{\label{step:1 load}\load($data$, $\ket{\psi}$)}
  {
  	$\ket{\psi_{0_0}} = \ket{u}\ket{m} =
  	\ket{1;0,\cdots,0}$ \\ \label{step:2 initial state CVO-QRAM}
  	\ForEach{$(x_k,p_k) \in \data$}{ \label{step:3 loop}
    	\label{rstep3}
  		\label{t} $t=$ The number of bits with value $ 1 $ in the pattern $ p_k $ \\
  	  \label{list} $l=$ a list containing the positions of $ p_k $ where $ p_k [j] =
  	  1.$ \\
  	  \BlankLine
  	  \BlankLine
   	 
  		$\ket{\psi_{k_1}} =\prod \limits_{l_i\in l} CX_{(u,m[l_i])}\ket{\psi_{k_0}}$ \\  \label{rstep6}
  		\BlankLine
  		\BlankLine
  		$\ket{\psi_{k_2}} = {{C^{t}U^{(x_k,\gamma_k)}}_{(m[l_0,l_1,\ldots l_{t-1}],u)}}\ket{\psi_{k_1}}$\\\label{rstep7}
  		\BlankLine
  		\BlankLine
  		\If{$k \neq M-1$}
  		{$\ket{\psi_{k_3}} =\prod \limits_{l_i\in l} CX_{(u,m[_i])}\ket{\psi_{k_2}}$ \\\label{rstep8}}
  	}
  	\KwRet\ $\ket{m}$
	}
	\caption{CVO-QRAM - Complex Data Storage Algorithm}
	\label{alg:CV-QRAM optimization}
\end{algorithm}

Define the amplitude remaining to be loaded 
by $\gamma_0=1$, $\gamma_{k+1}=\gamma_k-|x_{k+1}|^2$ for $0\leq k\leq M-1$. Notice that by normalization of the input data, $\gamma_{M-1}=0$. Then, for each $k$, define the unitary:
$$
U^{(x_k,\gamma_k)}=\frac{1}{\sqrt{\gamma_k}}\begin{pmatrix}
\sqrt{\gamma_k-|x_k|^2}&x_k\\
x_k^*&\sqrt{\gamma_k-|x_k|^2}
\end{pmatrix}.$$

The algorithm correctness is proven using the following loop invariant $H_k$, $k=0...M-1$:
$$
H_k:\text{Right before loading $(x_k,p_k)$, line 3, the state of the system is $\ket{\psi_{k_0}}=\sum_{j=0}^{k-1}x_j\ket{0,p_j}+\sqrt{\gamma_k}\ket{1,0}$}.$$
For $k=0$, $H_0$ holds vacuously. Let us check for $k<M-1$, the for loop maintains the property. Assume that $H_k$ is verified. Then, simply compute:
\begin{equation}
\begin{split}
\ket{\psi_{k_1}} &= \sum_{j=0}^{k-1}x_j\ket{0,p_j}+\sqrt{\gamma_k}\ket{1,p_k}\\
\ket{\psi_{k_2}} &= \sum_{j=0}^{k-1}x_j\ket{0,p_j}+\sqrt{\gamma_k}\left(\frac{x_k}{\sqrt{\gamma_k}}\ket{0,p_k}+\frac{\sqrt{\gamma_{k+1}}}{\sqrt{\gamma_k}}\ket{1,p_k}\right)\\
&= \sum_{j=0}^{k}x_j\ket{0,p_j}+\sqrt{\gamma_{k+1}}\ket{1,p_k}\\
\ket{\psi_{k_3}}&=\sum_{j=0}^{k}x_j\ket{0,p_j}+\sqrt{\gamma_{k+1}}\ket{1,0}
\end{split}
\end{equation}
Thus, the property is maintained. In order to show that the procedure ends well, simply notice $\ket{\gamma_{M-1}}=0$ since the input data must be normalized (without loss of generality).\\
The CVO-QRAM method needs two steps of classical pre-processing. First, given an input with $M$ patterns to load, the classical device must first sort the input patterns with cost $\mathcal O(M\log M)$. Then, it creates the circuits associated to the $U^{(x_k,\gamma_k)}$ with total cost $\mathcal O(nM)$. The total, classical cost is $\mathcal O(M\log M+nM)$. In terms of quantum gates, the number of CNOT gates to prepare a state is
$$
\sum_{t=1}^n\mu_t(8t-4)-t_{\text{max}}
,$$ 
where $\mu_t$ is the number of input patterns $p_k$ with $t$ bits with value $1$ in the binary string and $t_{\text{max}}$ is the highest value of $t$ with $\mu_t\neq0$. 
\end{appendix}
\section{Overlap-ADAPT-VQE }

The general workflow of the Overlap ADAPT-VQE algorithm is given in the following. The computations of overlap and gradients of the overlap are not expensive. We encourage the reader to consult our most recent publications\cite{Feniou_2023, feniou2023greedy} (and supplementary information) on the subject for further details about how the overlap and the overlap-gradients are computed in Overlap-ADAPT-VQE, or any other related technical aspect.

\begin{algorithm}
  \SetKwFunction{load}{Overlap-ADAPT-VQE}
  \SetKwInOut{Input}{input}\SetKwInOut{Output}{output}
  \Input{initial state $\ket{\psi_0}$, target state $\ket{\phi_\text{T}}$, pool of operators $\mathcal{A}$, convergence threshold $\epsilon$}
  \Output{ansatz $\ket{\psi(\theta)}$ such as ~$\lvert\braket{\psi(\theta)\lvert\phi_{\text{T}}}\rvert~<~\epsilon$}

    \BlankLine
  \Fn{\load{$\ket{\psi_{0}}$, $\ket{\phi_T}$, $\mathcal{A}$, $\epsilon$}}
  {
    $n \leftarrow 1$ \\
    \While{$S_n = \lvert\braket{\psi_n(\theta)\lvert\Psi_{T}}\rvert>\epsilon$}
    {
      \ForEach{$A_k \in \mathcal{A}$}
      {
        Measure gradients $g_{A_k} \leftarrow \left| \frac{\partial S_{n, A_k}(\theta)}{\partial \theta} \right| \bigg|_{\theta = 0}$ \\
        }
    $i \leftarrow$ index of the largest element of $\tilde{g}$ \\
    Pick best operator $A_n \leftarrow A_i$ \\
    Grow Ansatz $\ket{\psi_n} \leftarrow A_n\ket{\psi_{n-1}}$ \\
    \BlankLine
    Optimise Ansatz $\theta^{\rm{opt}} = \underset{\theta = (\theta_1, \ldots, \theta_{n})}{\operatorname{argmin}}~\lvert\braket{\psi_n(\theta)\lvert\Psi_{T}}\rvert$
    
    $\theta \leftarrow \theta^{opt}$\\
    Iterate $n \leftarrow n + 1$ \\
      }
    \KwRet\ $\ket{\psi_n(\theta)}$
  }
  \caption{Overlap-ADAPT-VQE - Variational Quantum State Preparation}
  \label{alg:Overlap-ADAPT-VQE}
  
\end{algorithm}

\subsection{Operator Pools}\label{sec:Pools}\vspace{2mm}

The literature contains a wide range of operator pools for ADAPT-like algorithms and VQEs. Some are chemically-inspired and ensure that the ansatz preserves the symmetry of the target state, while others prioritize hardware efficiency and minimize the number of operators used \cite{tang2021qubit, kandala2017hardware}. This section intends to offer a concise introduction to the two operator pools that have been used in prior numerical simulations.
\vspace{3mm}

\noindent \textbf{The Qubit Excitation-based Pool}\vspace{2mm}

The Qubit excitation-based (QEB) pool is widely used for simulating quantum chemical systems \cite{yordanov2021qubit} because it preserves spin and number of particles while still being fairly hardware efficient. The QEB pool consists of so-called single-qubit and double-qubit excitation operators which take the form

\begin{align}\label{eq:single}
A_{pq}=\frac{1}{2}\left(X_qY_p-Y_pX_q\right).
\end{align}
and
\begin{equation}\label{eq:double}
\begin{split}
A_{pqrs}&=\frac{1}{8}(X_rY_sX_pX_q+Y_rX_sX_pX_q+Y_rY_sY_pX_q +Y_rY_sX_pY_q\\&-X_rX_sY_pX_q-X_rX_sX_pY_q-Y_rX_sY_pY_q-X_rY_sY_pY_q).
\end{split}
\end{equation}

The variables $p, q, r, s$ represent qubit indices, and $X_p$ and $Y_p$ represent the standard one-qubit Pauli gates applied to qubit $p$. Therefore, the single-qubit generator $A_{pq}$ operates between individual qubits $p$ and $q$, while the double-qubit generator $A_{pqrs}$ operates between the qubit pairs $(p, q)$ and $(r, s)$. The computation of parametric exponentiation for QEB operators is straightforward, and the resulting unitary operators come with well established CNOT-optimised circuit implementations given below.

\begin{figure}[h]
\centering
\scalebox{0.9}{
\Qcircuit @C=1.0em @R=1em @!R { \\
	 	\nghost{{q}_{r} :  } & \lstick{{q}_{r} :  } & \gate{\mathrm{R_Z}\,(\mathrm{\frac{\pi}{2}})} & \qw & \ctrl{1} & \gate{\mathrm{R_Y}\,(\mathrm{\frac\theta2})} & \ctrl{1} & \gate{\mathrm{R_Y}\,(\mathrm{-\frac\theta2})} & \ctrl{1} & \qw & \qw\\
	 	\nghost{{q}_{s} :  } & \lstick{{q}_{s} :  } & \gate{\mathrm{R_Y}\,(\mathrm{-\frac{\pi}{2}})} & \gate{\mathrm{R_Z}\,(\mathrm{-\frac{\pi}{2}})} & \targ & \gate{\mathrm{R_Z}\,(\mathrm{-\frac{\pi}{2}})} & \targ & \gate{\mathrm{H}} & \targ & \qw & \qw}}
\caption{A quantum circuit performing a generic single-qubit excitation \cite{yordanov2020efficient}.}
\label{fig:sqe}
\end{figure}
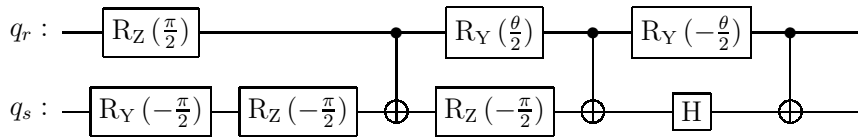

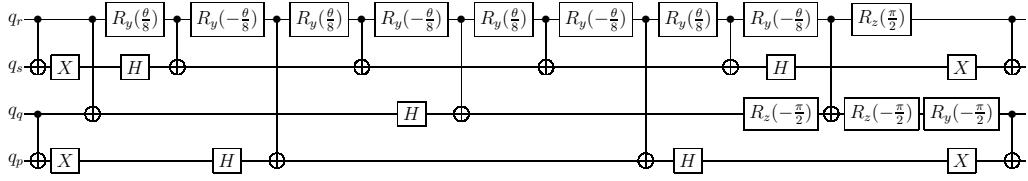
\begin{figure}[ht]
\centering
\scalebox{0.6}{
\Qcircuit @C=0.3em @R=1.0em {
&&& \\ 
q_r \ && \ctrl{1} & \qw & \ctrl{2} & \gate{R_y(\frac{\theta}{8})} & \ctrl{1} & \gate{R_y(-\frac{\theta}{8})} & \ctrl{3} & \gate{R_y(\frac{\theta}{8})} & \ctrl{1} & \gate{R_y(-\frac{\theta}{8})} & \ctrl{2} & \gate{R_y(\frac{\theta}{8})} & \ctrl{1} & \gate{R_y(-\frac{\theta}{8})} & \ctrl{3} & \gate{R_y(\frac{\theta}{8})} & \ctrl{1} & \gate{R_y(-\frac{\theta}{8})} & \ctrl{2} &  \gate{R_z(\frac{\pi}{2})} & \qw & \ctrl{1} & \qw \\
q_s \ && \targ{} & \gate{X} & \qw & \gate{H} & \targ & \qw & \qw & \qw & \targ & \qw & \qw & \qw & \targ & \qw & \qw & \qw & \targ & \gate{H} & \qw & \qw & \gate{X} & \targ & \qw & \\
q_q \ && \ctrl{1} & \qw & \targ{} & \qw & \qw  & \qw & \qw & \qw & \qw & \gate{H} & \targ & \qw & \qw  & \qw & \qw & \qw & \qw & \gate{R_z(-\frac{\pi}{2})} & \targ & \gate{R_z(-\frac{\pi}{2})}& \gate{R_y(-\frac{\pi}{2})}  & \ctrl{1} & \qw & \\
q_p \ && \targ{} & \gate{X} & \qw & \qw & \qw & \gate{H} & \targ & \qw & \qw & \qw & \qw & \qw & \qw & \qw & \targ & \gate{H} & \qw & \qw & \qw  & \qw &\gate{X} & \targ & \qw &  \\
}}
\caption{A quantum circuit performing a generic double-qubit excitation \cite{yordanov2020efficient}.}
\label{fig:d_q_exc_full}
\end{figure}

\vspace{5mm}

\noindent \textbf{The Qubit Hardware-efficient Pool }\vspace{2mm}

The Qubit hardware-efficient \cite{ryabinkin2018qubit} pool instead uses decomposed single and double excitation operators which are of the form
\begin{align}\label{eq:single_2}
B_{pq}=X_qY_p
\end{align}
and
\begin{equation}\label{eq:double_2}
\begin{split}
 B^{(1)}_{pqrs}=& X_rY_sX_pX_q, \quad B^{(2)}_{pqrs}= Y_rX_sX_pX_q, \quad B^{(3)}_{pqrs}= Y_rY_sY_pX_q, \quad B^{(4)}_{pqrs}= Y_rY_sX_pY_q,\\ 
 B^{(5)}_{pqrs}=& X_rX_sY_pX_q, \quad B^{(6)}_{pqrs}= X_rX_sX_pY_q, \quad B^{(7)}_{pqrs}= Y_rX_sY_pY_q, \quad B^{(8)}_{pqrs}= X_rY_sY_pY_q,
\end{split}
\end{equation}
where $p, q, r, s$ again denote qubit indices and $X_p$ and $Y_p$ are one-qubit Pauli gates acting on qubit $p$. Note that the qubit hardware-efficient pool does not preserve particle and spin conservation of the ansatz, which may result in convergence difficulties due to the violation of such crucial system symmetries. However, prior numerical simulations employing the qubit hardware-efficient pool have produced ansatze that require fewer CNOT operations for implementation on quantum hardware when compared to ansatz wave functions based on the QEB pool.
\subsection{Target states details}\label{sec:target}

\begin{table}[H]
    \footnotesize
    \centering
    \begin{tabular}{|c|c|c|c|}
        \hline
        Number of Determinants & \multicolumn{2}{c|}{Fidelity}\\
        \cline{2-3}
        & CIPSI iterate state & Truncated ground state \\
        \hline
        8 & 0.0397 & 0.0627 \\
        21 & 0.0562 & 0.1015 \\
        46 & 0.0761 & 0.1501 \\
        92 & 0.1103 & 0.2121 \\
        185 & 0.1495 & 0.2749 \\
        403 & 0.1824 & 0.3498 \\
        827 & 0.2147 & 0.4631 \\
        1657 & 0.2954 & 0.6164 \\
        3356 & 0.3905 & 0.7800 \\
        7054 & 0.5071 & 0.8833 \\
        14174 & 0.6411 & 0.9390 \\
        28976 & 0.8757 & 0.9932 \\
        58664 & 0.9964 & 1.0000 \\
        \hline
    \end{tabular}
    \caption{H$_{14}$ ground state approximations : comparative fidelities of variational CIPSI iterate states and truncations of the ground state with respect to the numbers of determinants.}
    \label{tab:mytable}
\end{table}
\begin{acknowledgement}
We thank Muhammad Hassan and Yvon Maday (LJLL, Sorbonne Université) for providing feedback on the composition of this paper. Computations have been performed at IDRIS (Jean Zay) on GENCI grant no A0150712052 (JPP). Additional DGX-A100 GPUs computations have been performed thanks to the support of the European Research Council (ERC) under the European Union's Horizon 2020 research and innovation program (grant agreement No 810367), project EMC2 (JPP). Support from the PEPR EPiQ and HQI programs is also acknowledged (JPP).
\end{acknowledgement}
\bibliography{achemso-demo}

\end{document}